\documentclass[a4paper]{article}
\usepackage{INTERSPEECH2021}
\usepackage{multirow}
\usepackage{tipa}
\usepackage{mwe}
\usepackage{subfigure}

\title{Acoustical Analysis of Speech Under Physical Stress in Relation to \\ Physical Activities and Physical Literacy}
\name{Si-Ioi Ng$^1$, Rui-Si Ma$^2$, Tan Lee$^1$, Raymond Kim-Wai Sum$^2$}
\address{
  $^1$Department of Electronic Engineering, 
  $^2$Department of Sports Science and Physical Education \\ The Chinese University of Hong Kong}
\email{siioing@link.cuhk.edu.hk, tanlee@ee.cuhk.edu.hk}

\begin{document}

\maketitle
\begin{abstract}

Human speech production encompasses physiological processes that naturally react to physic stress. Stress caused by physical activity (PA), e.g., running, may lead to significant changes in a person's speech. The major changes are related to the aspects of fundamental frequency level, speaking rate, pause pattern, and breathiness. The extent of changes depends presumably on the physical fitness and well-being of the person, as well as the intensity of PA. The general wellness of a person is further related to his/her physical literacy (PL), which refers to a holistic description of engagement in PA. This paper presents the development of a Cantonese speech database that contains audio recordings of speech before and after physical exercises of different intensity levels. The corpus design and data collection process are described. Preliminary results of acoustical analysis are presented to illustrate the impact of PA on F0 level, F0 range, speaking and articulation rate, and time duration of pauses. It is also noted that the effect of PA is correlated to some of the PA and PL measures.




\end{abstract}

\section{Introduction}\label{intro}
Speech, as one of the most natural and common media of human communication, carries a rich array of information that extends far beyond the verbal message being expressed. 
From an utterance, the speaker's gender, age, dialectal background, emotional status and personality could be identified by human listeners.
Part of the paralinguistic information in speech carries physiological and healthy condition of a speaker.
It is feasible to capture the natural speech sound in the form of acoustic signals, and extract the speaker's information via signal processing techniques and statistical modeling \cite{dehak2010front, mohammadi2010voice}.  
Acoustic and linguistic analysis of speech signals are effective means of detecting and quantifying disorders, diseases, and other changes in human body.


Speech production is an effortful activity that involves the brain and the respiratory system. It is closely coupled with breathing. 
While talking, inhalations are shortened and accelerated, while exhalations are prolonged and slowed down \cite{conrad1979speech}. 
The timing of inhalations depends on the control of habitual respiratory cycles \cite{trouvain2015prosodic}, and is mostly determined in accordance with the linguistic structure of speech \cite{rochet2013interplay}. 
If the functioning of a person's respiratory system does not function properly, 
evident changes in terms of perceptual and acoustic properties of the produced speech can be observed. 
In our daily lives, the pattern of breathing is often altered by physical activity (PA). 
PA refers to any body movement that increases the energy expenditure from resting state and reaches exercise state. 
It may take place during running, swimming or weight lifting, etc. \cite{kumar2015physical}. 
In exercise states, PA of different intensities pose different levels of physical stress to human body. Respiratory rate increases accordingly with the stress to provide adequate oxygen supply to human body. 
The effect of PA on breathing is expected to be manifested in the produced speech.
In \cite{godin2015physical}, it was shown that perceivable prosodic characteristic of speech, such as pause pattern, shifting of fundamental frequency (F0) and formants might be consistent indicators of physical stress. 
Detection of exercise intensity in PA was successfully performed using paralinguisic features of speech  \cite{truong15_interspeech}. 

The well-being of a person is related to participation in PA.
For older adults, regular physical exercises was found to improve subjective well-being measures, i.e.  happiness, satisfaction with life  \cite{mcauley2000social}. 
Adolescents who reported greater engagement in PA tend to experience less stress and depression \cite{norris1992effects}, and have higher self-perceived physical competence  \cite{timo2016perceived}, which is one’s ability to move with competence in a wide
variety of activities \cite{edwards2017definitions}. 
Recently, lights are shed on the concept of physical literacy (PL), which extends beyond the participation of PA \cite{whitehead2010physical}.
A physically literate individual should understand the value and responsibility, and is confident, motivated, skilful in participating in PA throughout the lifetime. 
Numerous instruments for assessing PL were proposed and validated \cite{tremblay2010physical}. Positive correlation between perceived PL and PA was reported in a large population study \cite{choi2018relationship}.

PL is correlated with PA, while PA determines the well-being and competence of a person, and affects the characteristics of breathing and speaking. Intuitively, a physically literate person with strong competence tends to have good respiratory control. This hypothetically implies less noticeable effect of physical stress, in terms of the difference in speech characteristics between resting and exercise states.
The present study starts with the collection of speech data in PA of different intensities. The speech database is intended to support investigation on the relationship between PA/PL and speech; facilitate statistical analysis of speech features and physiological measures; provide training data for classification of PA intensity experienced by the subjects.
The design of PA follows the Canadian Assessment of Physical Literacy-2 (CAPL-2) test protocol \cite{longmuir2018canadian}.
For general population, speech production may become laborious and difficult during and after PA. This results in breathy, intermittent and unstable voice as compared to that produced in resting state. 
A fundamental hypothesis of this research is that the influence of PA on speech production can be revealed by characteristic changes of acoustic speech signals. 
F0 and temporal characteristics are extracted from the speech signals to reflect the difference of speech between resting and exercise states. The correlation between the difference, and the measures of PA and PL, is investigated. 


\section{Design of Speech Corpus}

\subsection{About Cantonese}
\begin{figure}[h!]
  \setlength\belowcaptionskip{-0.8\baselineskip}
  \centering

  \includegraphics[width=\linewidth]{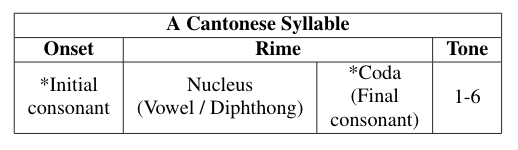}
  \caption{Structure of Cantonese syllable. '*' denotes optional.} 
  \label{fig:syllable_structure}
\end{figure}

Cantonese is a major Chinese dialect widely spoken in Hong Kong, Macau, Guangdong and Guangxi Provinces of Mainland China as well as overseas Chinese communities. 
Each Chinese character is pronounced as a single syllable carrying a lexical tone. 
The lexical tone (T) is determined by the temporal pattern of fundamental frequency (F0) across the voiced portion of a syllable. Cantonese has six tones, including T1  (high-level), T2 (high-rising), T3 (mid-level),  T4 (low-falling), T5 (low-rising)  and  T6  (low-level) \cite{bauer2011modern}.  
As shown in Figure \ref{fig:syllable_structure}, a legitimate Cantonese syllable can be divided into an onset and a rime. 
Onset is a consonant, while rime can contain either a nucleus or a nucleus followed by a coda. Nucleus can be a vowel or a diphthong, and coda can be a final consonant. 
Both onset and coda are optional in a Cantonese syllable. 
In the inventory of Cantonese phonemes, there are a total of $19$ initial consonants, $11$ vowels, $11$ diphthongs and $6$ final consonants \cite{bauer2011modern,lee2002spoken}. 

\subsection{Design of speech material}
The speech materials consist of three parts. 
The first part contains two text passages. Passage A is the story of `the North Wind and the Sun'. It comprises 142 syllables. The story content was originated from \cite{zee1991chinese}. It was used in the studies of Cantonese perceptual evaluation of voice \cite{law2010construction} and speech rhythm \cite{mok08_speechprosody}. 
Passage B is composed by following the corpus designed in \cite{truong15_interspeech} to 
elicit designated types of syntactic breaks and investigate how placing of pauses interacts with breathing.
The second part of speech materials contains three sustained vowels, namely the 
low-central vowel /\textipa{a:}/, high-front vowel /\textipa{i:}/ and mid-back vowel /\textipa{O:}/. Sustained vowels are most useful to reflect voice characteristics in stable articulation states. 
The third part contains digit sequences ``3947'', ``3948'', and ``0526''.
The digits cover all of the 6 Cantonese tones. The digit sequences are intended for reliably measuring F0 level and F0 range of a Cantonese speaker.

\subsection{Recording}

Speech recording took place in an outdoor sports ground of basketball court size. 
Under the instruction of qualified fitness trainers, each subject was required to go through three different parts of PA from CAPL-2. The first part was the Canadian Agility and Movement Skill Assessment (CAMSA) \cite{longmuir2017canadian}. It measured the skill execution and completion time for a series of dynamic movement to assess the subject's movement capabilities. The second exercise was plank, a reliable measure of isometric core muscular endurance \cite{schellenberg2007clinical}. The third part was the Progressive Aerobic Cardiovascular Endurance Run (PACER). It assessed aerobic endurance and determined the maximal oxygen consumption ($\text{VO}_{2}$ max) of the subjects \cite{scott2013ability}. 
The physical intensity was presumably increasing from CAMSA to PACER. Each subject was required to wear heart rate monitors. The average heart rate during speaking was regarded as an objective indicator of the PA intensity \cite{karvonen1988heart}.

Speech recordings were made before and in-between the physical exercises. 
Since the recordings shared the same linguistic content, they were most suitable for contrastive investigation on the influence of PA on paralinguistic factors of speech.
A digital recorder (TASCOM DR-44WL) was located at $20$-$50$ centimeters in front of the participant’s mouth. 
The gain of the recorder was adjusted to maintain the background noise level below -$30$ dB (relative to the maximum input level).  
A two-channel stereo recording was made for each subject at a sampling rate of 44.1 kHz.
Each recording was later down-sampled to 16 kHz with mono-channel, and was further divided into three sections corresponding to the three parts of text materials. 

\subsection{Subjects and measurement of PA and PL}
30 native Cantonese speakers (23 males, 7 females), aged between 18 to 21, were recruited by convenience sampling. All speakers were university students majored in sports science. 
The Physical Activity Readiness Questionnaires (PAR-Q) were distributed to all subjects \cite{shephard1988q}. All of them indicated ``NO'' to all questions, and became eligible to participate in the physical exercises. Informed consents were obtained 
prior to the recording session. The subjects were required to complete the questionnaire of the 9-item Perceived Physical Literacy Instrument (PPLI) \cite{sum2018perceived}. The responses were rated on 5-point Likert scale ranging from “strongly disagree” to “strongly agree”. The 9 items, denoted as PL1-9, considered the “knowledge and understanding”, “self-expression and communication with others”, and “sense of self and self-confidence” of a person.
The subjects were also asked to report their daily routine on PA in sedentary, light, moderate and vigorous intensities.  
Metabolic equivalent of task (MET), which represents the amount of PA in daily life, was measured based on the reported routine \cite{fan2014chinese}. One MET is defined as an oxygen uptake of 3.5 ml per kilogram per minute while a person is sitting at rest. MET minutes per week was calculated for each level of PA intensity. 
The subject's performance in terms of time and skill scores in CAMSA, total elapsed time in plank, and number of round finished in PACER, was measured. 

\section{Speech Feature Extraction}

\subsection{Extraction of F0}
F0 in a person's speech carries linguistic, paralinguistic and physiological information. 
F0 was extracted using the auto-correlation (AC) function via Praat \cite{jadoul2018introducing, boersma1993accurate}. 
Reliability of F0 estimation is sensitive to the parameter settings, specifically, the floor ($F0_{floor}$) and ceiling values ($F0_{floor}$). Extracting the F0 for different speakers within the same floor and ceiling values may not be optimal.  
Male voice typically has lower F0 than female. F0 range is also related to language and emotional status of the speaker \cite{traunmuller1995frequency}. 
We expect that physical stress would cause large alteration of F0. 
A two-stage strategy of F0 extraction was adapted to improve the accuracy of F0 estimation \cite{evanini2010importance}. 
In the first-pass extraction, the floor and ceiling values were initialized to 75 Hz and 600 Hz. Subsequently the 35-th quantile ($q_{35}$) and the 65-th quantile ($q_{65}$) of the F0 values were obtained. 
Before the second-pass processing, the floor and ceiling values were updated as $q_{35} * 0.72 - 10$ and $q_{65} * 1.9 + 10$. These equations were manually tuned to better correlate between automatic and manual estimation of F0 \cite{de2009automatic}. 
The F0 contour was re-estimated as the final output. The utterance-level F0 was obtained by averaging the F0 values of the vowel segments.
Average F0 values of tone 1 (T1), tone 3 (T3) and tone 6 (T6) in each utterance were calculated. They represented the high, medium and low level of F0 in the utterance.

We are interested in how the F0 level changes from resting state to exercise states. The change of F0 was measured in terms of cents (\textcent), defined as,
\begin{equation*}
\text{\textcent} = 1200 * \log_{2}{\frac{F0_{e}}{F0_{r}}}
\end{equation*}
The variation of the F0 contour within an utterance is represented by dynamism quotient (DQ), which is given by dividing the standard deviation by mean \cite{traunmuller1995frequency, hincks04_icall}. Higher DQ indicates larger variation of F0. 
The change of DQ from resting state to exercise states was measured as $\text{DQ}_{e}/\text{DQ}_{r}$.

\begin{table}[t!]
\caption{Mean (standard deviation) of speech features.}
\centering
\resizebox{0.99\linewidth}{!}{
\begin{tabular}{|c|c|c|c|c|c|}
\hline
\multirow{2}{*}{\textbf{Feature}}                                         & \multirow{2}{*}{\textbf{Measure}} & \multirow{2}{*}{\textbf{Rest}} & \multirow{2}{*}{\textbf{CAMSA}} & \multirow{2}{*}{\textbf{Plank}} & \multirow{2}{*}{\textbf{PACER}} \\
                                                                          &                                  &                                &                                 &                                 &                                 \\ \hline
\multirow{4}{*}{\begin{tabular}[c]{@{}c@{}}F0\\ (utterance)\end{tabular}} & \multirow{2}{*}{Hz}              & \multirow{2}{*}{143 (38)}      & \multirow{2}{*}{147 (41)}       & \multirow{2}{*}{144 (40)}       & \multirow{2}{*}{159 (41)}       \\
                                                                          &                                  &                                &                                 &                                 &                                 \\ \cline{2-6} 
                                                                          & \multirow{2}{*}{DQ (\%)}         & \multirow{2}{*}{15.58 (2.90)}  & \multirow{2}{*}{16.12 (3.10)}   & \multirow{2}{*}{15.95 (3.34)}   & \multirow{2}{*}{16.48 (2.65)}   \\
                                                                          &                                  &                                &                                 &                                 &                                 \\ \hline
\multirow{4}{*}{\begin{tabular}[c]{@{}c@{}}F0\\ (Tone 1)\end{tabular}}    & \multirow{2}{*}{Hz}              & \multirow{2}{*}{162 (45)}      & \multirow{2}{*}{166 (48)}       & \multirow{2}{*}{163 (46)}       & \multirow{2}{*}{178 (46)}       \\
                                                                          &                                  &                                &                                 &                                 &                                 \\ \cline{2-6} 
                                                                          & \multirow{2}{*}{DQ (\%)}         & \multirow{2}{*}{14.29 (3.05)}  & \multirow{2}{*}{15.01 (3.15)}   & \multirow{2}{*}{15.22 (3.35)}   & \multirow{2}{*}{15.63 (3.06)}   \\
                                                                          &                                  &                                &                                 &                                 &                                 \\ \hline
\multirow{4}{*}{\begin{tabular}[c]{@{}c@{}}F0\\ (Tone 3)\end{tabular}}    & \multirow{2}{*}{Hz}              & \multirow{2}{*}{142 (39)}      & \multirow{2}{*}{144 (41)}       & \multirow{2}{*}{143 (40)}       & \multirow{2}{*}{157 (40)}       \\
                                                                          &                                  &                                &                                 &                                 &                                 \\ \cline{2-6} 
                                                                          & \multirow{2}{*}{DQ (\%)}         & \multirow{2}{*}{11.45 (2.73)}  & \multirow{2}{*}{12.41 (2.90)}   & \multirow{2}{*}{11.79 (2.70)}   & \multirow{2}{*}{12.69 (2.3)}    \\
                                                                          &                                  &                                &                                 &                                 &                                 \\ \hline
\multirow{4}{*}{\begin{tabular}[c]{@{}c@{}}F0\\ (Tone 6)\end{tabular}}    & \multirow{2}{*}{Hz}              & \multirow{2}{*}{133 (37)}      & \multirow{2}{*}{137 (40)}       & \multirow{2}{*}{135 (39)}       & \multirow{2}{*}{147 (39)}       \\
                                                                          &                                  &                                &                                 &                                 &                                 \\ \cline{2-6} 
                                                                          & \multirow{2}{*}{DQ (\%)}         & \multirow{2}{*}{10.95 (2.55)}  & \multirow{2}{*}{11.30 (2.31)}   & \multirow{2}{*}{12.18 (3.28)}   & \multirow{2}{*}{12.60 (2.47)}   \\
                                                                          &                                  &                                &                                 &                                 &                                 \\ \hline
\begin{tabular}[c]{@{}c@{}}Speaking \\ rate\end{tabular}                  & \multirow{2}{*}{spm}             & 246 (34)                       & 267 (38)                        & 267 (34)                        & 252 (37)                        \\ \cline{1-1} \cline{3-6} 
\begin{tabular}[c]{@{}c@{}}Articulation \\ rate\end{tabular}              &                                  & 306 (36)                       & 341 (41)                        & 340 (42)                        & 366 (36)                        \\ \hline
\begin{tabular}[c]{@{}c@{}}Speech-to-pause \\ time ratio\end{tabular}   & Ratio                            & 4.33 (1.25)                    & 3.90 (1.25)                     & 4.02 (1.35)                     & 2.35 (0.73)                     \\ \hline
\end{tabular}
}
\label{pitch}
\end{table}

\subsection{Timing characteristics} 
Increased respiratory rate is a typical response to physical stress. 
To understand how speech properties change with intensity level of physical stress, temporal characteristics of speech including speaking rate, articulation rate and speech-to-pause time ratio were examined. 
Speaking rate is defined as the number of elicited syllables per minute (spm). Articulation rate has the same definition as speaking rate, except that the time intervals occupied by non-speech segments are excluded. 
Speech-to-pause time ratio is the total time of speech divided by the total duration of silence. Beside syntactic pauses, breathing pauses are frequently found in speech under physical stress. 
The speech-to-pause ratio is expected to carry the information of breathing strategy in speech under physical stress.
The change of speech-to-pause time ratio (tr) from resting state to exercise states is expressed by $\text{tr}_{e} / \text{tr}_{r}$. 
For speaking rate (sr) and articulation rate (ar), the difference between the two states are calculated as by $\text{sr}_{e} - \text{sr}_{r}$, and $\text{ar}_{e} - \text{ar}_{r}$, respectively.

To automatically locate the time boundaries of phones and syllables, 
forced alignment was applied to the utterances with the Gaussian Mixture Model-Hidden Markov Model (GMM-HMM) tri-phone acoustic model. We were focused on the passage of ``The North Wind and The Sun". 
The acoustic model was trained on a large-vocabulary database of adult speech called CUSENT. The corpus contained about 20 hours of speech recorded from $76$ adult speakers \cite{lee2002spoken}. The acoustic features for GMM-HMM training consisted of $13$-dimensional Mel-frequency cepstral coefficients (MFCC) and their first- and second-order derivatives extracted at every $0.010$ second. Feature space Maximum Likelihood Linear Regression (fMLLR) was applied to the acoustic features\cite{gales1998maximum}. Forced alignment and acoustic model training were implemented by the Kaldi speech recognition toolkit \cite{povey2011kaldi}.

\begin{table}[t!]
\caption{Difference of speech features between resting and exercise states.}
\centering
\resizebox{0.95\linewidth}{!}{
\begin{tabular}{|c|c|c|c|c|}
\hline
\multirow{2}{*}{\textbf{Feature}}                                         & \multirow{2}{*}{\textbf{\begin{tabular}[c]{@{}c@{}}Measure of \\ Difference\end{tabular}}} & \multirow{2}{*}{\textbf{Rest-CAMSA}} & \multirow{2}{*}{\textbf{Rest-Plank}} & \multirow{2}{*}{\textbf{Rest-PACER}} \\
                                                                          &                                                                                            &                                      &                                      &                                      \\ \hline
\multirow{4}{*}{\begin{tabular}[c]{@{}c@{}}F0\\ (utterance)\end{tabular}} & \multirow{2}{*}{Cents}                                                                     & \multirow{2}{*}{39 (84)}             & \multirow{2}{*}{11 (81)}             & \multirow{2}{*}{181 (141)}           \\
                                                                          &                                                                                            &                                      &                                      &                                      \\ \cline{2-5} 
                                                                          & \multirow{2}{*}{DQ ratio (\%)}                                                             & \multirow{2}{*}{104.30 (13.40)}      & \multirow{2}{*}{102 (12.37)}         & \multirow{2}{*}{107.15 (14.98)}      \\
                                                                          &                                                                                            &                                      &                                      &                                      \\ \hline
\multirow{4}{*}{\begin{tabular}[c]{@{}c@{}}F0\\ (Tone 1)\end{tabular}}    & \multirow{2}{*}{Cents}                                                                     & \multirow{2}{*}{37 (99)}             & \multirow{2}{*}{0 (103)}             & \multirow{2}{*}{168 (153)}           \\
                                                                          &                                                                                            &                                      &                                      &                                      \\ \cline{2-5} 
                                                                          & \multirow{2}{*}{DQ ratio (\%)}                                                             & \multirow{2}{*}{105.96 (13.35)}      & \multirow{2}{*}{107.10 (13.55)}      & \multirow{2}{*}{110.99 (17.44)}      \\
                                                                          &                                                                                            &                                      &                                      &                                      \\ \hline
\multirow{4}{*}{\begin{tabular}[c]{@{}c@{}}F0\\ (Tone 3)\end{tabular}}    & \multirow{2}{*}{Cents}                                                                     & \multirow{2}{*}{29 (95)}             & \multirow{2}{*}{10 (86)}             & \multirow{2}{*}{181 (159)}           \\
                                                                          &                                                                                            &                                      &                                      &                                      \\ \cline{2-5} 
                                                                          & \multirow{2}{*}{DQ ratio (\%)}                                                             & \multirow{2}{*}{112.55 (30.89)}      & \multirow{2}{*}{105.63 (21.49)}      & \multirow{2}{*}{115.12 (26.88)}      \\
                                                                          &                                                                                            &                                      &                                      &                                      \\ \hline
\multirow{4}{*}{\begin{tabular}[c]{@{}c@{}}F0\\ (Tone 6)\end{tabular}}    & \multirow{2}{*}{Cents}                                                                     & \multirow{2}{*}{33 (80)}             & \multirow{2}{*}{16 (77)}             & \multirow{2}{*}{172 (128)}           \\
                                                                          &                                                                                            &                                      &                                      &                                      \\ \cline{2-5} 
                                                                          & \multirow{2}{*}{DQ ratio (\%)}                                                             & \multirow{2}{*}{105.52 (20.13)}      & \multirow{2}{*}{112.46 (21.77)}      & \multirow{2}{*}{119.23 (29.09)}      \\
                                                                          &                                                                                            &                                      &                                      &                                      \\ \hline
\begin{tabular}[c]{@{}c@{}}Speaking \\ rate\end{tabular}                  & \multirow{2}{*}{\begin{tabular}[c]{@{}c@{}}spm\\ difference\end{tabular}}                  & 21 (29)                              & 21 (33)                              & 6 (37)                               \\ \cline{1-1} \cline{3-5} 
\begin{tabular}[c]{@{}c@{}}Articulation \\ rate\end{tabular}              &                                                                                            & 35 (27)                              & 34 (33)                              & 61 (34)                              \\ \hline
\begin{tabular}[c]{@{}c@{}}Speech-to-pause \\ time ratio\end{tabular}   & Ratio (\%)                                                                                 & 91.76 (24.12)                        & 94.59 (24.40)                        & 56.79 (19.49)                        \\ \hline
\end{tabular}
}
\label{feature_difference}
\end{table}

\section{Results and Discussion}

\begin{table*}[t!]
\caption{Correlation between the acoustic changes in exercise states and the PA/PL measures.}
\centering
\resizebox{0.75\textwidth}{!}{
\begin{tabular}{|c|c|c|c|c|c|c|c|}
\hline
\multirow{4}{*}{\textbf{Feature}}                                                     & \multirow{4}{*}{\textbf{Measure}}                                          & \multicolumn{2}{c|}{\multirow{3}{*}{\textbf{Rest-CAMSA}}} & \multicolumn{2}{c|}{\multirow{3}{*}{\textbf{Rest-Plank}}} & \multicolumn{2}{c|}{\multirow{3}{*}{\textbf{Rest-PACER}}} \\
                                                                                      &                                                                           & \multicolumn{2}{c|}{}                                     & \multicolumn{2}{c|}{}                                     & \multicolumn{2}{c|}{}                                     \\
                                                                                      &                                                                           & \multicolumn{2}{c|}{}                                     & \multicolumn{2}{c|}{}                                     & \multicolumn{2}{c|}{}                                     \\ \cline{3-8} 
                                                                                      &                                                                           & PA/PL Measures                   & PCC                          & PA/PL Measures                   & PCC                          & PA/PL Measures                       & PCC                      \\ \hline
\multirow{4}{*}{\begin{tabular}[c]{@{}c@{}}F0\\ (utterance)\end{tabular}}             & Cents                                                                     & -                          & -                            & -                          & -                            & -                              & -                        \\ \cline{2-8} 
                                                                                      & \multirow{3}{*}{DQ ratio}                                                 & PL1                        & -0.456                       & PL3                        & -0.377                       & \multirow{3}{*}{MET (total)}   & \multirow{3}{*}{0.420}   \\
                                                                                      &                                                                           & Pacer round                & -0.399                       & PL9                        & -0.377                       &                                &                          \\
                                                                                      &                                                                           & CAMSA skill score          & -0.501                       & MET (total)                & 0.379                        &                                &                          \\ \hline
\multirow{2}{*}{\begin{tabular}[c]{@{}c@{}}Speaking \\ rate\end{tabular}}             & \multirow{3}{*}{\begin{tabular}[c]{@{}c@{}}spm\\ difference\end{tabular}} & \multirow{2}{*}{PL1}       & \multirow{2}{*}{-0.410}      & \multirow{2}{*}{PL1}       & \multirow{2}{*}{-0.560}      & PL1                            & -0.540                   \\
                                                                                      &                                                                           &                            &                              &                            &                              & CAMSA time score               & -0.376                   \\ \cline{1-1} \cline{3-8} 
\begin{tabular}[c]{@{}c@{}}Articulation \\ rate\end{tabular}                          &                                                                           & PL1                        & -0.403                       & PL1                        & -0.466                       & PL1                            & -0.465                   \\ \hline
\multirow{2}{*}{\begin{tabular}[c]{@{}c@{}}Speech-to-pause\\ time ratio\end{tabular}} & \multirow{2}{*}{Ratio}                                                    & PL2                        & -0.388                       & PL1                        & -0.513                       & PPLI                           & -0.376                   \\
                                                                                      &                                                                           & Pacer round                & 0.374                        & PL2                        & -0.448                       & CAMSA time score               & -0.370                   \\ \hline
\end{tabular}
}
\label{feature_correlation}
\end{table*}

\subsection{Analysis of speech features}
Table \ref{pitch} reports the mean and the standard deviation of F0, speaking rate, articulation rate and speech-to-pause time ratio in resting and exercise states. 
The difference of feature values between resting and exercise states are shown in Table \ref{feature_difference}. 
Given limited amount of speech recordings, the gender effect is not considered in the analysis. 
ANOVA and the two-tailed t-test are applied. A significant threshold of 0.05 is assumed.

The results of ANOVA indicate the changes of F0 levels in different exercises are significantly different ($p<0.001$). An increase of over 150 cents is observed in PACER. 
The DQs in CAMSA, plank and PACER do not differ significantly. For each type of exercise, significant difference is found in the DQs of T1, T3, and T6 ($p<0.001$). Physical stress exhibits different extents of influence to different tones.

The two-tailed t-test shows that the speaking rate, articulation rate and speech-to-pause time ratio of exercise states are significantly different from that in resting state ($p<0.01$). ANOVA indicates the increase of speaking rate and articulation rate, and the decrease of speech-to-pause time ratio, are significantly different among CAMSA, plank and PACER ($p<0.001$). 
The temporal characteristics of speech in PACER are further compared to resting state via the t-test. 
A notable increase in articulation rate is observed ($p<0.001$), but not in speaking rate. Subjects generally requires more frequent and deeper breathing in high-intensity exercise. 
The speech-to-pause time ratio is lower in the exercise states ($p<0.001$).
These observations suggest that the subjects tend to talk faster and use longer time to breath in PACER. 
The above results can help with the design of acoustic feature sets that distinguish between PA of different intensities.

\subsection{Correlation with PA and PL measures}

We attempt to correlate the measures of PA/PL with the changes of acoustical/temporal characteristics of speech under physical stress.
The daily exercise routine of the subjects and their performance in different exercises are provided as the PA/PL measures. 
The correlation is measured by the Pearson's correlation coefficient (PCC). 
PA/PL measures that are significantly correlated to the changes of speech features are reported in Table \ref{feature_correlation}.

The changes of speaking rate and articulation rate in exercise states are correlated mostly to PL1 and PL2,  
which correspond to `I am physically fit in accordance with my age' and 'I have a positive attitude and interest in sports' respectively. 
As subjects are confident with their physical fitness and enthusiastic in sports, the temporal characteristics of speech are less altered during exercises.  
The self-perceived physical fitness in youngsters is a reliable predictor of actual fitness \cite{germain2006relationship}. 
Subjects who rated themselves physically fit could be in fact in good shape, and better control their breathing during exercises. This leads to similar speaking and articulation rates between resting and exercise states. 
For speech-to-pause time ratio, subjects who perform well in CAMSA, i.e. high CAMSA time score, have less changes in the time ratio in  PACER. Subjects who endured more rounds in PACER enlarges the speech-to-pause time ratio more in CAMSA. 
A negative correlation is found in between the PPLI score and the speech-to-time ratio in PACER. The overall descriptor of a person's PL is connected to characteristic changes of speech.

PA/PL measures do not significantly correlate to the shift of F0 in exercise states ($p>0.05$). 
Changes of DQ have negative correlation with PL1, PL3 and PL9. The descriptions of PL3 and PL 9 are `I appreciate myself or others doing sports' and `I am aware of the benefits of sports related to health'. Both PL3 and PL9 concern the knowledge and understanding of participating in PA.
A better understanding towards PA, and better performance in PA, i.e. higher CAMSA skill score and more PACER rounds, lead to less deviation of F0 contour during exercises. 
The frequency of engagement in PA also plays a significant role in the changes of acoustic features in exercise states. Higher total MET score corresponds to larger DQs. 

\section{Conclusion}
We present the collection of speech data for investigating the relation between physical stress and speech. Statistical analysis has shown that there are different extent of changes in fundamental frequency, speaking and articulation rate, and speech-to-pause time ratio in different physical activities.
Variations of F0 in different Cantonese tones are different in the resting and exercise states. Correlation analysis indicates that part of the measures of physical activities and physical literacy, and daily habits of participating in physical activities, are associated with the changes in acoustic/ temporal characteristics of speech.
To extend the present study, more speech data will be collected from subjects of diverse background. In-depth analysis of the 6 Cantonese tones under physical stress, investigation on the interaction between pauses and breathing, classification of physical stress, and automatic assessment of physical literacy will be among the future works.

\section{Acknowledgements}\label{ack}
\vspace{0.5mm}
This research is partially supported by a GRF project grant (Ref: CUHK 14208020) from Hong Kong Research Grants Council.

\bibliographystyle{IEEEtran}

\bibliography{mybib}


\end{document}